%% file: main.tex
\documentclass[sigconf]{acmart}
\usepackage{listings}
\usepackage{booktabs}
\usepackage[justification=centering]{subfig}
\usepackage{float}
\usepackage{caption}
\usepackage{balance}
\usepackage{colortbl}
\usepackage{adjustbox}
\usepackage{tablefootnote}
\usepackage[T1]{fontenc}
\usepackage[utf8]{inputenc}
\usepackage{multirow}
\usepackage{amsthm}
\usepackage{graphicx}

\copyrightyear{2021}
\acmYear{2021}
\setcopyright{acmcopyright}\acmConference[EASE 2021]{Evaluation and Assessment
in Software Engineering}{June 21--23, 2021}{Trondheim, Norway}
\acmBooktitle{Evaluation and Assessment in Software Engineering (EASE 2021), June
21--23, 2021, Trondheim, Norway}
\acmPrice{15.00}
\acmDOI{10.1145/3463274.3463337}
\acmISBN{978-1-4503-9053-8/21/06}

\begin{document}
\title{On the Nature of Issues in Five Open Source Microservices Systems: An Empirical Study}

\author{Muhammad Waseem$^{1}$, Peng Liang$^{1*}$, Mojtaba Shahin$^{2}$, Aakash Ahmad$^{3}$, Ali Rezaei Nasab$^{4}$}
\affiliation{%
  \institution{$^{1}$School of Computer Science, Wuhan University, Wuhan, China}
  \institution{$^{2}$Faculty of Information Technology, Monash University, Melbourne, Australia}
  \institution{$^{3}$College of Computer Science and Engineering, University of Ha'il, Ha'il, Saudi Arabia}
  \institution{$^{4}$Department of Engineering, Computer Science and Information Technology, Shiraz University, Shiraz, Iran}
  \institution{\{m.waseem, liangp\}@whu.edu.cn,  mojtaba.shahin@monash.edu,  a.abbasi@uoh.edu.sa,  rezaei.ali.nasab@gmail.com}
  \country{}
}

\renewcommand{\shortauthors}{M. Waseem et al.}

\begin{comment}
\author{Muhammad Waseem}
\affiliation{\institution{{\textit{School of Computer Science}\\
\textit{Wuhan University}\\
Wuhan, China\\
m.waseem@whu.edu.cn}}
}
             
\author{Peng Liang}
\affiliation{\institution{{\textit{School of Computer Science}\\
\textit{Wuhan University}\\
Wuhan, China\\
liangp@whu.edu.cn}}
}
\authornote{Corresponding author}
             
\author{Mojtaba Shahin}
\affiliation{\institution{{\textit{Faculty of Information Technology}\\
\textit{Monash University}\\
Melbourne, Australia\\
mojtaba.shahin@monash.edu}}
}

\author{Aakash Ahmad}
\affiliation{\institution{{\textit{College of Comput Sci. \& Eng.}\\
\textit{University of Ha'il}\\
Ha'il, Saudi Arabia\\
a.abbasi@uoh.edu.sa}}
}
             
\author{Ali Rezaei Nasab}
\affiliation{\institution{{\textit{Department of Engineering, Computer Science and Information Technology}\\
\textit{Shiraz University}\\
Shiraz, Iran\\
rezaei.ali.nasab@gmail.com}}
}
\end{comment}

\begin{abstract}
Due to its enormous benefits, the research and industry communities have shown an increasing interest in the Microservices Architecture (MSA) style over the last few years. Despite this, there is a limited evidence-based and thorough understanding of the types of issues (e.g., faults, errors, failures, mistakes) faced by microservices system developers and causes that trigger the issues. Such evidence-based understanding of issues and causes is vital for long-term, impactful, and quality research and practice in the MSA style. To that end, we conducted an empirical study on 1,345 issue discussions extracted from five open source microservices systems hosted on GitHub. Our analysis led to the first of its kind taxonomy of the types of issues in open source microservices systems, informing that the problems originating from Technical debt (321, 23.86\%), Build (145, 10.78\%), Security (137, 10.18\%), and Service execution and communication (119, 8.84\%) are prominent. We identified that “General programming errors”, “Poor security management”, “Invalid configuration and communication", and “Legacy versions, compatibility and dependency" are the predominant causes for the leading four issue categories. Study results streamline a taxonomy of issues, their mapping with underlying causes, and present empirical findings that could facilitate research and development on emerging and next-generation microservices systems.
%We recommend directions for developing microservices systems that require attention.

%wrong inheritance hierarchies and cyclic dependency (e.g., between microservices) are the dominant causes of Technical Debt. 
%Further, limiting the hierarchies to a single class and eliminating the dependency at the MSA design level are the prevalent solutions for code and service design debt, respectively.
%We discovered that wrong inheritance hierarchies and cyclic dependency (e.g., microservices) are the dominant causes for microservices technical debt, and limiting the hierarchies to a single class and eliminating dependency on MSA design level are the prevalent solutions. %Finally, our study concludes that X is needed to help microservices systems developers reduce X and Y.
\end{abstract}
\begin{CCSXML}
<ccs2012>
<concept>
<concept_id>10011007.10011074.10011075</concept_id>
<concept_desc>Software and its engineering~Designing software</concept_desc>
<concept_significance>500</concept_significance>
</concept>
</ccs2012>
\end{CCSXML}

\ccsdesc[500]{Software and its engineering~Designing software}
\ccsdesc[500]{General and reference~Empirical studies}
\keywords{Microservice, Open Source Software, Microservices Architecture, Issue, Empirical Study}

\maketitle
%\justifying
\input{introduction.tex}

\input{relatedwork.tex}

\input{researchmethod.tex}

\input{results.tex}

\input{discussion.tex}

\input{threats.tex}

\input{conclusions.tex}
%\input{ack.tex}

\begin{acks}
This work has been partially supported by the National Key R\&D Program of China with Grant No. 2018YFB1402800.
\end{acks}

\balance

\bibliographystyle{ACM-Reference-Format}
\bibliography{references}
\end{document}

%% file: introduction.tex
\section{Introduction}
\label{sec:introduction}

The software industry has recently witnessed a growing popularity of microservices architecture (MSA) style as a promising design approach to develop applications that consist of multiple small, manageable, and independently deployable services \cite{fowler2014microservices, dragoni2017microservices}. Software development organizations may have adopted or plan to use the MSA style for various reasons. Specifically, some want to increase the scalability of applications using the MSA style, while others use it to quickly release new products and services to the customer, whereas it is argued that the MSA style can also help build autonomous development teams \cite{taibi2017processes, jamshidi2018microservices}. From an architectural perspective, a microservices system (system that adopts the MSA style) entails a significant degree of complexity both at the design phase as well as at runtime configuration \cite{newman2020building}. This implies that the MSA style brings unique challenges for software organizations, and many quality attributes may be (positively or negatively) influenced \cite{dragoni2017microservices, waseem2020systematic}. For example, service level security may be impacted because microservices are developed and deployed by various technologies (e.g., Docker containers \cite{combe2016docker}) and tools that are potentially vulnerable to security attacks \cite{newman2020building, yu2019survey}. 
% In contrast to monolithic systems or traditional service-oriented systems, an acceptable level of observability in microservices systems is achieved by continuously monitoring and troubleshooting hundreds or even thousands of autonomous services running simultaneously~\cite{waseem2020systematic, newman2020building}.
Data management is also influenced because each microservice needs to own its domain data and logic \cite{wagner2018net}. This can, for example, challenge reaching consistency across multiple microservices.

Zimmermann argues that MSA is not entirely new from Service-Oriented Architecture (SOA) (e.g., \textit{``microservices constitute one particular implementation approach to SOA -- service development and deployment''}) \cite{zimmermann2017microservices}. Similarly, Márquez and Astudillo discovered that some existing design rationale and patterns from SOA hold for MSAs \cite{1-marquez2018actual}. However, an important body of literature (e.g., \cite{dragoni2017microservices, esposito2016challenges, jamshidi2018microservices, yarygina2018overcoming}) has concluded that there are overwhelming differences between microservices systems and monolithic systems and traditional service-oriented systems in terms of design, implementation, test, and deployment. Gupta and Palvankar indicated that even having SOA experience and background can lead to sub-optimal decisions (e.g., excessive service calls) in microservices systems development \cite{Dinkar2020Pitfalls}. Hence, microservices systems may have an \textit{additional} and \textit{specific} set of \textbf{issues}. Borrowing the idea from \cite{dragoni2017microservices, ren2020understanding}, we define \textbf{issues} in this study as errors, faults, failures, and bugs that occur in a microservices system and consequently impact its quality and functionality. Hence, there is a need for new practices, techniques, and tools or revising old ones to address the specific and additional issues for microservices systems.
%Furthermore, the specific and additional issues to microservices systems may need new and revised best practices and techniques that may not be found in other types of systems.  %Furthermore, the best practices and lessons learned from other types of systems may not help address issues in microservices systems.

A few studies have recently investigated \textit{particular} issues (e.g., performance issues) in microservices systems \cite{5-walker2020automated, 13-taibi2018definition, 7-zhou2018fault, 9-wu}. Despite these efforts, there is no in-depth and comprehensive study on the nature of different types of issues that microservices developers face and the potential causes of these issues. Jamshidi \textit{et al}. believe that this can be partially attributed to the fact that researchers have limited access to industry scale microservices systems \cite{jamshidi2018microservices}. The empirical knowledge on the nature of issues occurring in microservices systems can be useful from the following perspectives: (i) experienced microservices developers can be allocated to address the most frequent and challenging issues, (ii) novice microservices developers can quickly be informed of empirically-justified issues and avoid common mistakes, and (iii) the industry and academic communities can invest resources to develop tools and techniques for the frequently reported issues in microservices systems.
\textbf{This work aims to} systematically and comprehensively study and classify the issues that developers face in developing microservices systems and the causes that trigger these issues. To this end, we conducted an empirical study on 1,345 issue discussions collected from five Open Source Software (OSS) projects in GitHub. We started with identifying five OSS projects in GitHub designed by following the MSA style (open source microservices systems). Then, we extracted and manually analyzed 1,345 issues from these five open source microservices systems. 
%We focused on issue discussions on GitHub instead of Stack Overflow (SO) discussions because GitHub presents more developers' perspectives, while SO data is more from users' perspectives~\cite{han2020programmers}

 \textbf{Our key findings} are based on an issue taxonomy that indicates: (1) the issues in open-source microservices systems are classified into 17 categories, in which Technical debt (321, 23.86\%), Build (145, 10.78\%), Security (137, 10.18\%), and Service execution and communication (119, 8.84\%) issues are most frequently reported; and (2) the leading \textbf{causes} behind these issues are “General programming errors”, “Poor security management”, “Invalid configuration and communication”, and “Legacy versions, compatibility and dependency”.

%are the are only limited to specific issues, and developers are reluctant to discuss explicitly the reasons behind most of the issues occurring in microservices systems;% and (3) most of the \textbf{solutions} consist of pull request calls to address specific issue.%\textbf{the solutions of issues} are collected ... \textcolor{blue}{@Waseem - complete this part}.
%Our study found that the issues are mostly addressed by inviting proper pull requests. 

 \textbf{Our core contributions} are: (1) we develop the first of its kind taxonomy of issues occurring in open-source microservices systems based on a qualitative analysis of 1,345 issue discussions among developers, (2) we identify a comprehensive list of causes for identified issues and map them to identified issues, (3) we provide several potential research topics on microservices systems that require more attention, and (4) we publicly release a dataset to enable researchers and practitioners to access all collected data and replicate and validate our study~\cite{replpack}.

\textbf{Paper organization}: Section \ref{sec:relatedwork} reviews the related work. Section \ref{sec:researchmethod} explains our research method. Section \ref{sec:results} presents the results followed by a discussion on the key findings in Section \ref{sec:discussion}. Section \ref{sec:threats} reports validity threats. Section \ref{sec:conclusions} concludes the paper.

%% file: relatedwork.tex
\section{Related Work}
\label{sec:relatedwork}
\subsection{Mining Knowledge from Open-Source Microservices Systems}
\label{sec:miningfrommsa}
\textbf{Architectural patterns and tactics for MSA}: Architectural patterns and tactics represent generic and reuse-driven knowledge that enables developers and architects to rely on best practices during design and development of MSA-based systems~\cite{3-marquez2018empirical}. Several studies, such as \cite{1-marquez2018actual, 12-muntonimining}, focused on mining software repositories (e.g., analyzing version control, change logs, and design documents) to conduct postmortem analysis of historical data of open-source MSA-based systems. The postmortem analysis of repositories such as GitHub or logs of source code supports the empirical discovery of patterns to address various reuse-driven development \cite{1-marquez2018actual, 3-marquez2018empirical} and evolution issues \cite{10-balalaie2018microservices} of MSAs. Specifically, the studies \cite{1-marquez2018actual, 3-marquez2018empirical} investigated artifacts of MSA-based projects deployed on GitHub, such as developers’ discussion and source code files, to discover 17 architectural patterns that address reusability \cite{1-marquez2018actual} and 5 architectural tactics that resolve scalability issues of open-source MSA-based systems. In contrast to the design and development phase, the study in \cite{10-balalaie2018microservices} investigated the modernization processes of legacy software to discover and document 15 patterns for architecture-centric refactoring and migration of MSA-based systems. To complement repository mining approaches \cite{1-marquez2018actual, 3-marquez2018empirical}, the study in \cite{15-bandeira2019we} focuses on analyzing developers’ discussion – investigating human-centric view – regarding MSA development on SO to analyze the topics (e.g., processes, tools, issues, solutions, etc.) that developers find most exciting or challenging.

\textbf{Survey and analysis for issues of service evolvability}: In addition to repository mining approaches, several studies have focused on practitioners’ perspectives and system analysis to address issues specific to the maintenance and evolvability of MSA. For example, Bogner \textit{et al}. \cite{4-bogner2019assuring} conducted 17 structured interviews involving 14 practitioners from 10 MSA-based projects to investigate the role of tools, metrics, and patterns that manage issues like technical debt and enhance software evolvability. The surveyed practitioners recommended many techniques, such as code review or service cutting to address issues like granularity, composition, coupling, and service cohesion in MSA-based applications. Analysis of MSA-based systems, i.e., investigating source code and monitoring application execution, helps to identify the issues that could hinder or enhance the evolvability of microservices.

\subsection{Detecting Code Smells and Performance Issues in Microservices Systems}
\label{sec:issuesinmsa}
\textbf{Architectural smells for MSA}: Code smells, also referred to as architectural smells in MSA-based systems, indicate issues of poor coding and development practices that hinder source code understandability and ultimately impacts service maintenance and evolution \cite{13-taibi2018definition, 14-pigazzini2020towards}. The research in \cite{13-taibi2018definition} represents an empirical study that interviews 72 MSA developers to develop a catalog of 11 micro service-specific architectural smells that are frequently considered harmful by practitioners. For example, an architectural smell named \textit{‘Wrong Cuts’} corresponds to a bad practice of microservices being split based on technical layers (presentation, business, and data layers) instead of business capabilities. The wrong cuts for MSA result in a wrong separation of concerns that can increase data splitting and computation tasks for microservices. The solution presented in \cite{14-pigazzini2020towards} complements survey findings on identifying architectural smells in a semi-automated way by detecting three bad smells from 5 MSA-based systems. The studies \cite{13-taibi2018definition, 14-pigazzini2020towards} also pinpoint that in the absence of design strategies or architectural tactics, code smells could lead to functional and non-functional faults that impact the functionality and quality of MSA.

\textbf{Faults and performance issues in MSA}: A recently conducted industrial study in \cite{7-zhou2018fault} identifies and analyzes typical faults of microservices systems, the current practice of service debugging, and the challenges faced by developers in practice. For example, a fault such as \textit{‘transactional service failure’} is caused due to overloaded requests to a third-party (payment gateway) service, ultimately leading to a denial of service issues. In an attempt towards systematically addressing MSA faults, the authors in \cite{8-marquez2019identifying} examined the source code and documentation of 17 open source microservices systems to identify and document 5 architectural tactics of service availability. The study systematically identifies and characterizes architectural tactics in microservices-based systems to prevent faults relating to service availability and integrity. Other studies, such as \cite{2-cito2017empirical, 9-wu}, considered performance issues such as response time, throughput, and efficiency as runtime faults for MSA-based systems. For example, the study \cite{9-wu} contextualizes some faults from existing literature and datasets to present a solution named MicroService Root Cause Analysis (MicroRCA). MicroRCA identifies the root causes of performance-related faults by considering factors like resource utilization and throughput of the services.

\subsection{Conclusive Summary} 
\label{sec:relatedsummary} To conclude, we categorize the most relevant existing research as (a) repository mining approaches such as \cite{1-marquez2018actual, 2-cito2017empirical,3-marquez2018empirical} and (b) survey-based studies including \cite{4-bogner2019assuring, 7-zhou2018fault,13-taibi2018definition} to investigate MSA-based systems and their underlying issues. Specifically, the repository mining approaches exploit GitHub data (e.g., source code, design documents, docker files, and configuration scripts) to pinpoint a number of issues related to reusability and architectural quality \cite{1-marquez2018actual}, service scalability \cite{3-marquez2018empirical}, and evolvability \cite{2-cito2017empirical}. In comparison, our proposed study analyzes the real issues and their discussion by MSA developers on GitHub, to investigate \textit{what} are the issues and \textit{why} such issues are caused throughout the life-cycle of microservices systems. On the other hand, existing survey-based studies primarily focus on engaging MSA practitioners via structured questionnaire to identify issues related to service evolvability \cite{4-bogner2019assuring}, system faults \cite{7-zhou2018fault}, and bad smells \cite{13-taibi2018definition}. Such survey-based studies provide insights about various issues in the form of practitioners' (e.g., MSA architects) opinions and recommendations. 
In contrast to the existing survey-based studies (e.g., \cite{4-bogner2019assuring, 7-zhou2018fault,13-taibi2018definition}), our study leverages repository mining process to (i) extract issues, and (ii) taxonomically classify the issues with their root-cause analysis.%  our study presents the (i) comprehensive set of issues in terms of the taxonomy of issues, and (ii) list of causes, and their mapping with issues.

%% file: researchmethod.tex
\section{Research Method}
\label{sec:researchmethod}
%The research method to conduct this study comprises three phases, each detailed below and illustrated in Fig. \ref{fig:researchmethod}.

\begin{figure*}[!htbp]
  \centering
  \includegraphics[scale=0.51]{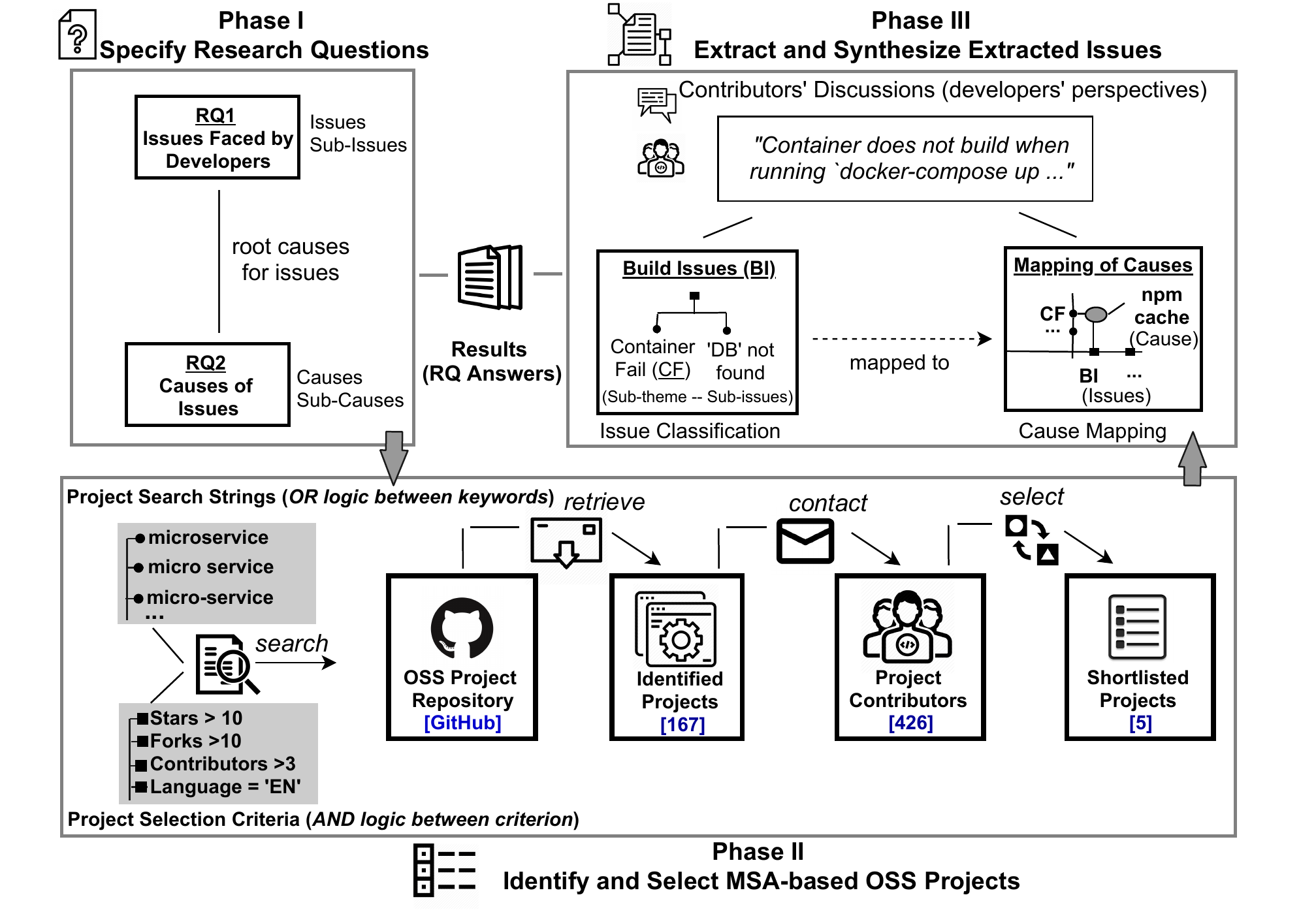}
  \caption{Overview of the research method}
  \label{fig:researchmethod}
\end{figure*}

\subsection{Phase I - Specify Research Questions (RQs)}
\label{Research Questions (RQs)}
We formulated two RQs that %streamline the scope and primary objectives of this study. These RQs also 
guide subsequent phases of the methodology as shown in Fig. \ref{fig:researchmethod}.

\textbf{RQ1}: \textit{What types of issues do developers face in open-source microservices systems?}

\textbf{Objective}: To systematically identify and taxonomically classify the types of issues that are faced by developers and impact open source microservices systems.
%\textbf{RQ2}: \textit{How long does it take to fix different types of issues in open-source microservices systems?}
%\textit{What is the difficulty levels of issues that occur in open-source microservices systems?}
%\\\textcolor{red}{\textbf{Objective(s)}: To pinpoint most challenging issues, such as 'service security', based on a number of criteria (e.g., time to resolve, number of contributors, contributor remarks, etc.)}\\

\textbf{RQ2}: \textit{What are the causes of issues that occur in open-source microservices systems?}

\textbf{Objective}: To investigate the causes and map them with identified issues that represent root causes of identified issues. %failures.
%to be resolved.

%\textbf{RQ3}: \textit{What solutions are proposed to fix issues that occur in open-source microservices systems?}

%\textbf{Objective}: To discover and document solutions as generic and repeatable strategies (e.g., \textit{`pick a set of standard terminologies -- project'} reducing technical debt) to resolve the identified issues.

\subsection{Phase II - Identify and Select Open-Source MSA Projects}
Identifying open-source MSA projects includes two steps: Keyword-based Search and Confirming with Contributors.

\textbf{Keyword-based Search}: To search the open source microservices systems in GitHub, Phase II in Fig. \ref{fig:researchmethod}, we followed the guidelines of systematic search from \cite{SystematicSearchMap2018} to compose search strings with format [\textit{keyword-1} <OR logic> \textit{keyword-2}, \ldots, \textit{keyword-N}], where keywords = [\textit{‘micro service’}, \textit{‘micro-service’}, \textit{‘microservice’}, \textit{‘Micro service’}, \textit{‘Micro-service’}, \textit{‘Microservice’}]. The search strings were searched on the title and the description of the projects on the GHTorrent dump hosted on Google Cloud version 1/4/2018. The search string returned 2,690 repositories. We then applied multi-criteria filtering with reference to \cite{GitHubFilter2020}, focused on popularity or perceived significance in the community (total stars), adoption by or interests of developers (total forks), and the number of developers (total contributors) for the projects \cite{Star-GitHub2018}.  We selected those repositories whose stars and forks were more than 10, the language was English, and contributors were more than 3. This led to 167 potential open source microservices projects.

\textbf{Confirming with Contributors}: We further confirmed with the core contributors whether these 167 projects follow the MSA style by asking them three questions:
%{We needed to eliminate the instances of false positive search results, such as misleading project names, mock-up code, and having only the “microservices” term in the title or description of a project.} %Hence, we needed to make sure if those 167 projects follow the MSA style. Therefore, to validate the identified projects, \textcolor{blue}{To this end, we contacted core contributors via their publicly available email IDs. We consider the top three contributors as core contributors who have had the commits in the projects, and their email IDs are publicly available. We got the 426 email IDs from 167 projects, and we emailed and asked them to answer the following three inquiry questions:}

(1) Can you please confirm if our interpretation is correct that this project (URL of the project) is designed by following the MSA style? Note that if the projects (e.g., frameworks or tools) support the development of microservices systems but are not designed based on the MSA style, please clarify it.

(2) If our interpretation is correct, what features or characteristics of the architecture of this open-source project show that the MSA style has been used?

(3) Optional question: Do you know any other open-source projects that are designed by following the MSA style?
 
We considered the top three contributors as core contributors of a project who have the most commits, and their email addresses are publicly available. Only 39 core contributors (9\% of 426 core contributors) responded to our inquiry. Based on the core contributors’ responses, we eventually concluded that five projects are designed based on the MSA style, and we selected them for data collection and synthesis. The synopsis of these five projects is presented in Table \ref{tab:selectedProjects}.

\begin{table}[H]
\caption{List of Identified open source microservices systems}\label{tab:selectedProjects}
\begin{adjustbox}{width=\columnwidth,center}
\begin{tabular}{lcccc}
\hline
\textbf{Project Name} & \textbf{Closed Issue} & \textbf{Contributors} & \textbf{Forks} & \textbf{Stars} \\ \hline
goa\tablefootnote{\url{https://github.com/goadesign/goa}}                & 799          & 82           & 447   & 4K    \\
eShopOnContainers\tablefootnote{\url{https://github.com/dotnet-architecture/eShopOnContainers}}  & 904          & 119          & 6.7K  & 16K   \\
light-4j\tablefootnote{\url{https://github.com/networknt/light-4j}}           & 523          & 28           & 492   & 2.9K  \\
Moleculer\tablefootnote{\url{https://github.com/moleculerjs/moleculer}}         & 436          & 76           & 377   & 3.9K  \\
Microservices-demo\tablefootnote{\url{https://github.com/microservices-demo/microservices-demo}} &280          & 45           & 1.5K  & 2.5K  \\ \hline
\end{tabular}
\end{adjustbox}
\end{table}

\subsection{Phase III- Extract and Synthesize Issues}
%\subsection{Phase III- Pilot study, Issues selection, and Data synthesize}
%\textbf{Pilot study}: Before extracting the actual data to answer our RQs, we conducted a pilot study to identify issues type, causes, and solutions. We randomly selected 200 issues from 2,942 issues from the 5 projects listed in Table \ref{tab:selectedProjects}. The contributor’s discussion about 200 issues was thoroughly investigated to understand the type of issue, causes, and solutions for open source microservices systems. In the end, we found that 120 issues can be used to answer our RQs.

\textbf{Issues screening}: %We selected 1,345 issues for their classification after thoroughly analyzing a total of 2,942 initially extracted issues. We excluded all those issues that consist of (i) general questions, opinions, feedback, and ideas, (ii) feature requests (e.g., enhancements, proposals), (iii) announcements, (iv) duplicated issues, and (v) issues without detailed description.
After selecting five open-source microservices systems, the first author manually retrieved the background information (e.g., issue label, URL) about issues of each system (see Table \ref{dataitems}). Only closed issues potentially provide answers to questions about the types (RQ1) and causes (RQ2). Therefore, we only extracted the information about closed issues. The total number of initially retrieved issues was 2,942. During this step, the first author thoroughly analyzed each of the 2,942 issues and excluded all those issues that consist of (i) general questions, opinions, feedback, and ideas, (ii) feature requests (e.g., enhancements, proposals), (iii) announcements, (iv) duplicated issues, and (v) issues without detailed description. There were several issues that the first author could not decide whether or not to include for further analysis. In such a situation, the first author discussed those issues with the second and third authors for their opinions about inclusion or exclusion. Any disagreements about the results of screening issues were discussed among all the authors to get a consensus. After completion of this step, we got 1,345 issues.
\begin{table}[H]
\caption{Data items extracted}\label{dataitems}
\begin{adjustbox}{width=\columnwidth,center}
\begin{tabular}{llp{6cm}}
\hline 
\textbf{\#}       & \textbf{Data item} & \textbf{Description} \\ \hline
D1 & Index  & The ID of the issue \\
D2 & Issue title  & The title of the issue \\
D3 & Issue link  & The weblink of the issue \\
D4 & Number of participants  & The number of practitioners who participate in the discussion \\
D5 & Issue (key points) & The identified key point(s) of the issue from practitioners’ discussion \\
D6 & Issue subcategory & The derived subcategory for the issue \\
D7 & Issue category & The derived category for the issue \\
D8 & Cause (key points) & The identified key point(s) for the cause from practitioners’ discussion \\
D9 & Cause subcategory  & The derived subcategory for the cause \\
D10 & Cause category & The derived category for the cause \\ \hline
\end{tabular}
\end{adjustbox}
\end{table}

\textbf{Data extraction}: To answer the RQs formulated in Section \ref{Research Questions (RQs)}, we defined a set of data items (see Table \ref{dataitems}) to be extracted from the discussion of selected issues. To check the viability of the defined data items, the first author conducted a pilot data extraction with 200 issues, and the rest of the authors evaluated the extracted data. After that, the first author used a revised set of data items for formal data extraction from the selected issues. All the authors then checked the extracted data to reduce potential bias and ambiguity. Data items (D1-D4) are used to provide the issue ID, issue title, issue link, and a who participated in the discussion about issues, whereas data items (D5-D10) are used to extract the data to answer the RQ1 and RQ2. Finally, we used spreadsheets to structure the data for further computation and synthesis of the extracted data.

\textbf{Data synthesis}: First, we applied thematic analysis \cite{cruzes2011recommended} to identify the categories of the issues, which is a method for systematically identifying, organizing, and offering insight into meaning (e.g., categories) of a dataset. It is composed of five steps. (i) Familiarizing with data: The first author repeatedly read the contributors discussion about issues and documented all points regarding issues and their causes discussed by the contributors. (ii) Generating initial codes: after data familiarization, first author produced an initial list of codes from the extracted data about microservices issues and their causes (see sheet Extracted Data in the dataset \cite{replpack}). (iii) Searching for the types of issues: We analyzed the initially generated codes and brought them under the specific types of issues. For instance, ``Code Smell and service dependency'' (see Fig. \ref{fig:Taxonomy}). (iv) Reviewing types of issues: all the authors reviewed and refined the coding results with the corresponding types of issues. We separated, merged, and dropped several issues based on mutual discussion between all authors during this step. (v) Defining and naming categories: During this step, we defined and further refined all the types of issues under precise and clear subcategories and categories (see Fig. \ref{fig:Taxonomy}). We introduced two levels of categories for managing the issues identified. First, we defined the subcategories to organize the types of issues under the specific subcategory (e.g., code debt for code smell). Then we arranged the subcategories under the precise categories (e.g., Code Debt in Technical debt). Furthermore, we also applied thematic analysis to identify, analyze, and classify the causes of issues (RQ2).

%% file: results.tex
\section{Results}
\label{sec:results}
This section reports the study results through analyzing and synthesizing the extracted data from the issues. We report the types of issues in Section \ref{sec:results_RQ1} and the causes of the leading issue categories in Section \ref{sec:results_RQ2}. The results of the issue types and causes are organized into three levels, including categories (e.g., \textbf{Technical debt}), subcategories (e.g., \textit{code debt}), and types (e.g., \textsc{code smell}). We present categories in \textbf{boldface}, subcategories in \textit{italic}, and types in \textsc{small capitals}.

\subsection{Types of Issues (RQ1)}
\label{sec:results_RQ1}
\begin{figure*}[!htbp]

    \flushleft
    \includegraphics[width=7.3in]{{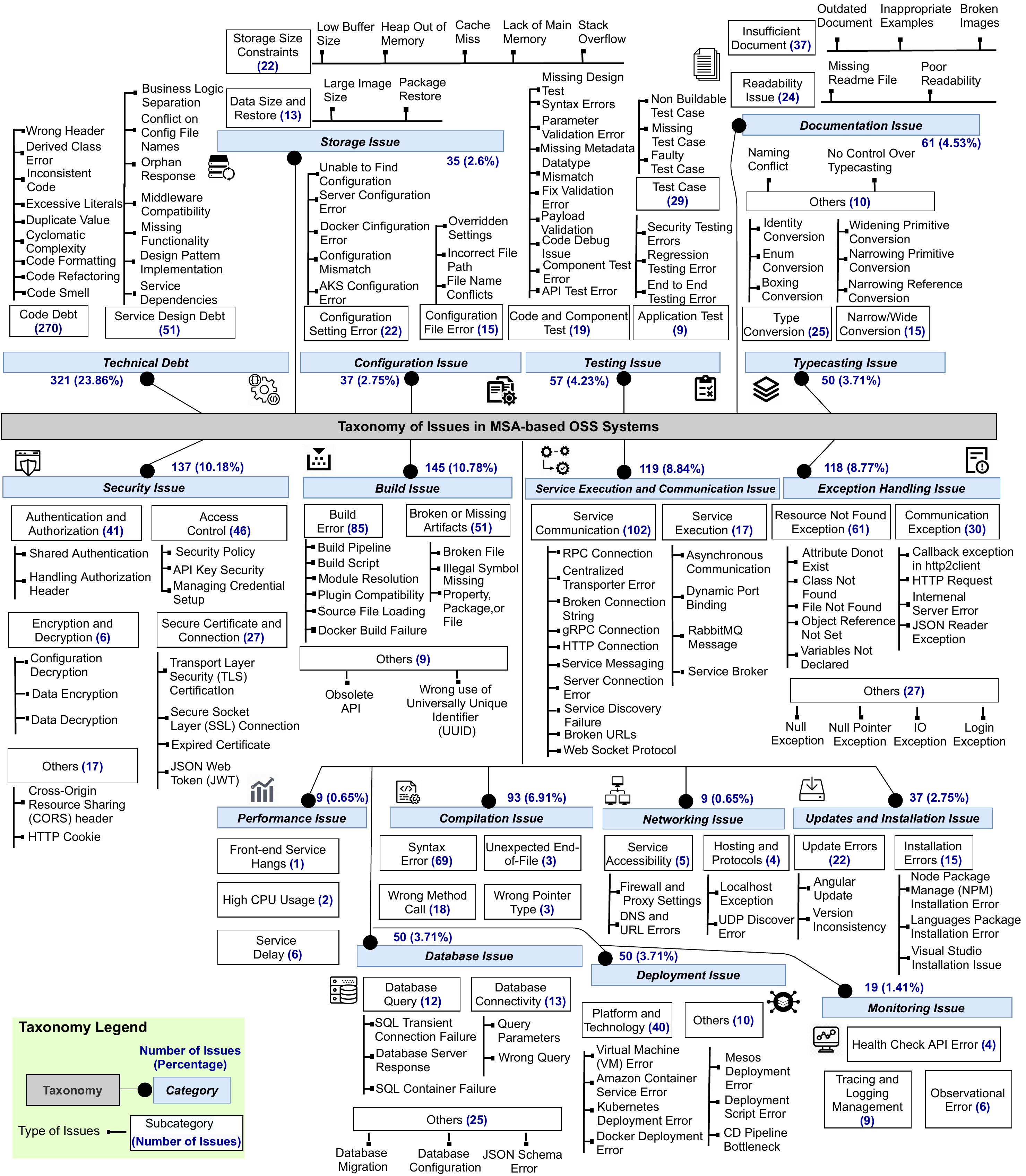}}    
        \caption{A taxonomy of issues in five open source microservices systems}
    \label{fig:Taxonomy}
\end{figure*}

Fig. \ref{fig:Taxonomy} presents the taxonomy of issues in open-source microservices systems. The results show that Technical debt (23.86\%), Build (10.78\%), Security (10.18\%), and Service execution and communication (8.84\%) issues are most frequently discussed. Number and percentage of the issues are shown around the corners of each category.
Due to the space limit, we only present the top four issue categories, and the complete list of categories, subcategories, and types of issues can be found in the dataset~\cite{replpack}. We provide examples in \textit{italic phrase} with the “\#” sign, which denotes the “issue number” in the “Master data sheet” in the dataset.

\textbf{\underline{Finding 1}}: We derived a taxonomy of open source microservices systems issues consisting of 17 categories, 46 subcategories, and 138 types, indicating the diversity of the issues in microservices systems.

%\begin{landscape}

\subsubsection{\textbf{Technical debt (TD)}} is identified as a type of critical issues in software development~\cite{li2015systematic}. We found 321 out of 1345 (23.86\%) issues related to \textit{code debt} and \textit{service design debt}. Below is an overview of each TD subcategory.

\textit{Code debt} refers to the source code’s issues, which could adversely impact the code’s quality. Most of the issues are related to \textsc{code smells}, \textsc{code refactoring}, and \textsc{code formatting}. For instance, goa developers found \textsc{code smells} in which ``\textit{attributes does not inherit properties if a list of attributes is specified in the Body function, \#143}''. Similarly, developers refactored the code of Microservices-demo project by “\textit{adding more domain logic and rules to the “Order Aggregate-Root” module and its methods, \#11}”. In addition, \textsc{derived class errors} (e.g., “\textit{property changes on derived classes do not occur in the mobile application, \#267}”), \textsc{cyclomatic complexity} (e.g., “\textit{goagen rendering code does not handle the cyclic dependency and infinity loops properly, \#260}”), and \textsc{wrong headers} (e.g., “\textit{using dial configuration header instead of Dial peer header in generated code, \#255}”) also negatively affect the code legibility of microservices systems.

%\textbf{Code debt} refers to the source code’s issues, which could adversely impact the code’s quality. Most of the issues are related to code smells, code refactoring, and code formatting. For instance, Goa developers found code smells in which “\textit{attributes does not inherit properties if a list of attributes is specified in the Body function, \#1097}”. Similarly, developers refactored the code of Microservices-demo project by “\textit{adding more domain logic and rules to the “Order Aggregate-Root” module and its methods, \#951}”. In addition, derived class errors (e.g., “\textit{property changes on derived classes do not occur in the mobile application, \#1234}”), Cyclomatic complexity (e.g., “\textit{goagen rendering code does not handle the cyclic dependency and infinity loops properly, \#1206}”), and wrong headers (e.g., “\textit{using Dial configuration header instead of Dial peer header in generated code, \#1252}”) also negatively affect the code legibility of microservices systems.

\textit{Service design debt} refers to the violation of adopting successful practices (e.g., MSA patterns) for designing open source microservices systems. We identified several types of issues related to \textit{service design debt} that can be further classified into \textsc{business logic separation} (e.g., “\textit{separation of business logic from user registration, \#245}”), \textsc{service dependencies} (e.g., “\textit{tests of service is failed in case of service have dependencies, \#290}”), and \textsc{missing functionality} (e.g., “\textit{missing proper file upload format to a generated client, \#281}”). Besides, we also observed \textsc{design patterns implementation} (e.g., “\textit{current interface does not support the implementation of multi-part/data decoder pattern, \#276}”), and \textit{orphan responses} (e.g., “\textit{orphan response is received, \#282}”) as service design debt.

\textbf{\underline{Finding 2}}: Most (i.e., 23.86\%, 321 of 1345) of open source microservices systems issues are related to \textbf{technical debt}, including \textit{Code debt} (e.g., \textsc{code smell}) and \textit{service design debt} (e.g., \textsc{service dependencies}).

\subsubsection{\textbf{Build issue}} Build is a process in which source code is converted into executable files (e.g., Jar, War, Ear, or Apk) for staging and production environments. We found that 10.75\% (i.e., 145 out of 1,345) issues are related to build issues, which are further classified into \textit{build errors}, \textit{broken and missing artifacts}, and \textit{others} subcategories (see Fig. \ref{fig:Taxonomy}). %  Further classification of these issues shows that most of build issues are related to Build script errors, Docker build fail, Missing properties, packages, and files. The other prominent subcategories are Plugin compatible errors, Source file loading errors, and Module resolution errors.

\textit{Build errors}: This category reports the errors related to \textsc{build script}, \textsc{Docker build failure}, \textsc{build pipeline}, \textsc{module resolution}, \textsc{plugin compatibility}, and \textsc{source file loading}. 

A build script is an input artifact to build systems written in the scripting language and carry the developer’s instruction to perform different build tasks. We identified several issues in which developers discuss \textsc{build script errors} (e.g.,“\textit{failed at the node-sass\@4.5.0 postinstall script ‘node scripts/build.js, \#366}”). The Docker build command builds Docker images from a Docker file and a “context”, which is a set of files located in the specified PATH or URL. The \textsc{Docker build failure} occurs when any context files are missing or configuration settings are incorrect. Developers received errors like “\textit{build error while connecting to Docker, \#431}”. Another type of build issue that we identified is regarding \textsc{module resolution errors} in which project modules are not correctly recognized or reloaded in the build system or different operating systems. For example, developers of Moleculer project discussed “\textit{Build module will not be reloaded when mod.js is changed, \#391}”, and developers of the light-4j project mentioned “\textit{config module build failing on windows environment, \#397}”. We also identified that developers reported a small ratio (i.e., 0.8\% issues) of \textsc{source file loading errors}. These errors mainly occurred when build systems are unable to load specified files for compilation. For instance, Microservices-demo developers were “\textit{unable to load the service index file for the compilation, \#425}”. It is common to use many third-party plugins or packages in open source microservices systems, and \textit{build errors} occur when existing code or used plugins are not compatible with each other. Our results indicate that developers face \textsc{plugin compatibility errors} during the build process of microservices systems. For instance, Microservices-demo developers mentioned this type of error “\textit{NU1202 Package Newtonsoft.Json 11.0.2 is not compatible with netcoreapp 2.0, \#398}”.

\textit{Broken and missing artifacts}: This subcategory covers the issues related to \textsc{missing artifacts} (i.e., properties, packages, files), \textsc{illegal symbols}, and \textsc{broken files}. This kind of error usually occurs in the build process’s parsing stage, in which build systems verify the required information in the build script before executing build tasks. \textsc{missing properties} or \textsc{file errors} occur when the required files or packages are not placed on a designated location or not imported into the required directory. For example, developers discussed that “\textit{Catalog.WebForms missing almost all files cant be compiled, \#319}” for Microservices-demo build. We identified several posts regarding \textsc{illegal symbols} found during the build process, which were triggered when the build system fails to resolve characters in the build script (e.g., “\textit{typo in the code, \#342}”, “\textit{invalid attribute type, \#390}”). We identified \textsc{broken file} issues that generally occur during the assembly stage of the build process. The assembly stage will not proceed further when there are duplicated or broken files. For instance, the Moleculer project’s build is failed “\textit{when KubeFed event broadcast is broken, \#293}”.

%\textbf{Others}: This subcategory covers the “Docker build fail”, “Obsolete API”, and “Wrong use of UUID” issues type.  We also identified errors regarding the use of obsolete APIs that are not used in the latest development branch. For instance, “\textit{Service ordering.api failed to build, \#160}”, an API used in the old version of the Microservices-demo project. Similarly, we also found several issues in which developers discuss “wrong use of UUID” packages (e.g.,“\textit{Build fails when UUID is used as an action parameter, \#163}”).

\textbf{\underline{Finding 3}}: Issues appearing in the build stage of open source microservices systems account for 10.78\% (145 of 1345) of the issues, mainly related to \textsc{build errors}, \textsc{Docker build failure}, and \textsc{broken or missing artifacts}, and \textsc{obsolete APIs}.

\subsubsection{\textbf{Security issue}} 
Microservices provide public interfaces, use network-exposed APIs for communicating with other services, and are developed by using polyglot technologies and toolsets that may be insecure. This makes microservices a potential target for cyber-attacks; therefore, security in open source microservices systems demands serious attention. Based on 132 Security issues, we define five subcategories (see Fig. \ref{fig:Taxonomy})

\textit{Authentication and authorization}: Authentication is the process of identifying a user, whereas authorization determines the access rights of the specified user to system resources. We found issues related to establishing secure \textit{authentication and authorization} mechanisms for open source microservices systems. Most of these subcategory issues are related to \textsc{handling authorization header} and
\textsc{shared authentication} issues. The issues associated with \textsc{handling authorization header} (e.g., Basic Auth, OAuth, OAuth 2.0) are generally used to implement authorization mechanisms. For instance, OAuth 2.0 lets the APIs authenticate and access the required resources. We found several issues related to Basic Auth, OAuth, or OAuth 2.0 header failure or non-availability. For example, the developers of goa project discussed the issue about “\textit{gRPC authentication fields missing, \#461}”, and the developers of the light-4j project mentioned that “\textit{OAuth helper needs to handle the error status from OAuth 2.0 provider, \#470}”. We also found many issues about an improper implementation of \textsc{shared authentication techniques}, e.g., “\textit{shared authentication methods are not generated, \#464}” in the goa project.

\textit{Security certificate and standard}: We found several issues about implementing \textit{security certificates and standards} for open source microservices systems. Security certificates (e.g., SSL, TSL) and standards (e.g., JWT) are used to secure connection and communication between two parties (e.g., client-server, service to service). For example, we found an issue “\textit{an SSL connection could not be established for Docker 2.3.0.3 with WSL2 support}” in the Microservices-demo project. Another issue is that “\textit{TLS certificates and OAuth2 certificates have conflicts, \#578}” over authorization and authentication in the light-4j project. Similarly, we found several issues in which the developers discussed errors, violations, and failures in implementing the \textsc{jwt token}. For instance, “\textit{JWT token fails, \#509}” in the light-4j project because light rest API expects JWT to be in the authorization module.

\textit{Encryption and decryption} refer to converting plain text into ciphertext and ciphertext into plain text to secure the information. We found several issues related to \textsc{data encryption}, \textsc{data decryption}, and \textsc{configuration decryption} of microservices systems. For example, the Microservices-demo project developers discussed that “\textit{The anti-forgery token could not be decrypted when the apps are hosted at the same location, \#494}”. As another example, “\textit{The values were not encrypted in a secret.yml file containing confidential information of the light-4j project, \#495}”.

We also found several Security issues about \textit{Access control} (e.g., \textsc{managing credential setups}, \textsc{API key security}) and \textit{Others} (e.g.,  \textsc{cors header}, \textsc{http cookies}). However, due to space limitation, we can not discuss all the \textbf{Security issues}. For details, see Fig. \ref{fig:Taxonomy} and the dataset \cite{replpack}.

\textbf{\underline{Finding 4}}: 10.18\% (135 of 1345) of the issues are identified as \textbf{Security issues} in open source microservices systems, mainly related to \textit{authentication and authorization}, \textit{access control}, \textit{encryption and decryption}, and \textit{secure certificate and connection}.

\subsubsection{\textbf{Service execution and communication issue}}
Communication problems are deceptive when services communicate across multiple servers and hosts in a distributed environment. Services interact using e.g., HTTP, AMQP, and TCP protocols depending on the nature of services. Our results show that 8.84\% (119 out of 1,345) issues are related to the execution and service communication of microservices with two subcategories.

\textit{Service communication}: This subcategory reports the issues, mainly related to \textsc{connection strings}, \textsc{communication protocols} (e.g., gRPC, HTTP, Web Socket, RPC), \textsc{centralized transporters} (e.g., MQTT, NATS, Kafka, Moleculer), \textsc{server connection}, \textsc{service discovery}, and \textsc{broken URLs}. The result reveals that most developers face \textsc{service discovery failure}, \textsc{HTTP protocol implementation}, and \textsc{centralized transporters} issues. For instance, API service discovery failed in light-4j project because of using “\textit{https connect to Web console, \#692}”.% in the result generated code for the “\textit{HTTP client can not distinguish between an empty slice and a nil slice}” for Goa project. %Similarly, a broker was failed to start with the MQTTs transporter in the moleculer project.

\textit{Service execution}: This subcategory reports the issues regarding \textsc{asynchronous communication}, \textsc{dynamic port binding}, \textsc{RabbitMQ messaging}, and \textsc{service broker} during service execution. These issues occur due to different reasons. For example, a dependency issue between microservices occurred when “\textit{integration commands were sent asynchronously, \#598}”. We also found several issues regarding \textsc{dynamic port binding}. For instance, a server module of light-4j project could not dynamically allocate a “\textit{port on the same host with a given range, \#616}”. Similarly, some developers face a situation in which “\textit{old services could not be replaced with new services because Service broker could not properly destroy the old services, \#690}”. 

\textbf{\underline{Finding 5}}: 8.84\% (119 out of 1345) of the issues are related to \textit{service execution} (e.g., \textsc{RPC connection}) and \textit{service communication} (e.g., \textsc{asynchronous communication}) of open source microservices systems.
\subsection{Causes of Issues (RQ2)}
\label{sec:results_RQ2}
We now report the causes of identified issues. For space reasons, we report and exemplify causes corresponding to the subcategories of the top four issue categories only. Illustrative details in Table 3 guide the interpretation for the remaining categories and subcategories of causes. To support such interpretations for extended details, a complete list of cause categories, cause subcategories, and types of causes can be found in the dataset \cite{replpack}. For the top four issue categories (see Fig.\ref{fig:Taxonomy}), we identified a total of 94 types of causes that can be classified into 7 categories and 21 subcategories.
%\textcolor{blue}{We derived 7 cause categories, 21 cause subcategories, and 94 types of causes for the top four issue categories. However, due to the space limit, we cannot present all these causes. The complete list of cause categories, cause subcategories, and types of causes can be found in the dataset \cite{replpack}. 
Overall, Table \ref{tab:issues-causes} maps the cause subcategories of leading issue categories (i.e., \textbf{TD}, \textbf{Build}, \textbf{Security}, \textbf{Service execution and communication}) to their cause categories, which are \textbf{General programming error (GPE)}, \textbf{Poor security management (PSM)}, \textbf{Invalid configuration and communication (ICC)}, \textbf{Legacy version, compatibility, and dependency problem (LCD)}. The other three categories are \textbf{Design level anomaly (DLA)}, \textbf{Missing features and artifacts (MFA)}, and \textbf{Insufficient resource (IR)}. The numbers (e.g., 158) in Table \ref{tab:issues-causes} denote the number of issues in a given issue subcategory that stems from a certain cause category. Note that not all the causes behind the issues are discussed. For instance, in the \textit{Others} subcategory of the \textbf{build issues}, we were not able to locate any cause for \textsc{obsolete API} and \textsc{wrong use of UUID} issues. Note that We identified 1007 (out of 1345) issues in which developers discuss the causes for the issues (see sheet Causes for All Issues in the dataset \cite{replpack}), 584 issues out of 1007 issues are belong to the leading four categories (see sheet Causes for Leading Issues in the dataset \cite{replpack}). Below, we report the top four cause categories with the issues caused by them.

\textbf{\underline{Finding 6}}: Overall, we derived 7 categories and 26 subcategories of causes from the 109 type of causes for all the issue categories, indicating the variety of the reasons behind issues in open source microservices systems.

\begin{table}[h]
\caption{Distribution of issue subcategories (vertical) and cause categories (horizontal)}
\label{tab:issues-causes}
\begin{adjustbox}{width=\columnwidth,center}
\begin{tabular}{lccccccc}
                                  & \multicolumn{1}{l}{\textbf{GPE}}     & \multicolumn{1}{l}{\textbf{PSM}} & \multicolumn{1}{l}{\textbf{ICC}} & \multicolumn{1}{l}{\textbf{LCD}} & \multicolumn{1}{l}{\textbf{DLA}} & \multicolumn{1}{l}{\textbf{MFA}} & \multicolumn{1}{l}{\textbf{IR}} \\
Code debt                         & \cellcolor[HTML]{4472C4}158 & \cellcolor[HTML]{F2F2F2}0        & \cellcolor[HTML]{F2F2F2}1          & \cellcolor[HTML]{F2F2F2}1          & \cellcolor[HTML]{8EAADB}33      & \cellcolor[HTML]{F2F2F2}9          & \cellcolor[HTML]{F2F2F2}0       \\
Service design debt               & \cellcolor[HTML]{F2F2F2}0   & \cellcolor[HTML]{F2F2F2}0        & \cellcolor[HTML]{F2F2F2}0          & \cellcolor[HTML]{F2F2F2}0          & \cellcolor[HTML]{D9E2F3}19      & \cellcolor[HTML]{F2F2F2}0          & \cellcolor[HTML]{F2F2F2}0       \\
Build errors                      & \cellcolor[HTML]{D9E2F3}18  & \cellcolor[HTML]{F2F2F2}0        & \cellcolor[HTML]{D9E2F3}12         & \cellcolor[HTML]{8EAADB}27         & \cellcolor[HTML]{D9E2F3}15      & \cellcolor[HTML]{F2F2F2}3          & \cellcolor[HTML]{D9E2F3}16      \\
Broken and missing artifacts      & \cellcolor[HTML]{F2F2F2}8   & \cellcolor[HTML]{F2F2F2}0        & \cellcolor[HTML]{F2F2F2}0          & \cellcolor[HTML]{F2F2F2}0          & \cellcolor[HTML]{F2F2F2}0       & \cellcolor[HTML]{8EAADB}31         & \cellcolor[HTML]{F2F2F2}0       \\
Build: Others                     & \cellcolor[HTML]{F2F2F2}0   & \cellcolor[HTML]{F2F2F2}0        & \cellcolor[HTML]{F2F2F2}0          & \cellcolor[HTML]{F2F2F2}0          & \cellcolor[HTML]{F2F2F2}0       & \cellcolor[HTML]{F2F2F2}0          & \cellcolor[HTML]{F2F2F2}0       \\
Access control                    & \cellcolor[HTML]{F2F2F2}1   & \cellcolor[HTML]{8EAADB}23       & \cellcolor[HTML]{D9E2F3}12         & \cellcolor[HTML]{F2F2F2}0          & \cellcolor[HTML]{F2F2F2}4       & \cellcolor[HTML]{F2F2F2}0          & \cellcolor[HTML]{F2F2F2}0       \\
Authentication and authorization: & \cellcolor[HTML]{F2F2F2}1   & \cellcolor[HTML]{D9E2F3}32       & \cellcolor[HTML]{F2F2F2}8          & \cellcolor[HTML]{F2F2F2}2          & \cellcolor[HTML]{F2F2F2}0       & \cellcolor[HTML]{F2F2F2}2          & \cellcolor[HTML]{F2F2F2}0       \\
Encryption and decryption         & \cellcolor[HTML]{F2F2F2}0   & \cellcolor[HTML]{F2F2F2}6        & \cellcolor[HTML]{F2F2F2}1          & \cellcolor[HTML]{F2F2F2}0          & \cellcolor[HTML]{F2F2F2}0       & \cellcolor[HTML]{F2F2F2}0          & \cellcolor[HTML]{F2F2F2}0       \\
Security certificate and standard & \cellcolor[HTML]{F2F2F2}0   & \cellcolor[HTML]{D9E2F3}23       & \cellcolor[HTML]{F2F2F2}1          & \cellcolor[HTML]{F2F2F2}5          & \cellcolor[HTML]{F2F2F2}0       & \cellcolor[HTML]{F2F2F2}0          & \cellcolor[HTML]{F2F2F2}0       \\
Security: Others                  & \cellcolor[HTML]{F2F2F2}0   & \cellcolor[HTML]{F2F2F2}8        & \cellcolor[HTML]{F2F2F2}1          & \cellcolor[HTML]{F2F2F2}0          & \cellcolor[HTML]{F2F2F2}0       & \cellcolor[HTML]{F2F2F2}1          & \cellcolor[HTML]{F2F2F2}0       \\
Service communication             & \cellcolor[HTML]{8EAADB}34  & \cellcolor[HTML]{F2F2F2}2        & \cellcolor[HTML]{8EAADB}34         & \cellcolor[HTML]{F2F2F2}6          & \cellcolor[HTML]{F2F2F2}4       & \cellcolor[HTML]{F2F2F2}0          & \cellcolor[HTML]{F2F2F2}3       \\
Service execution                 & \cellcolor[HTML]{F2F2F2}3   & \cellcolor[HTML]{F2F2F2}0        & \cellcolor[HTML]{D9E2F3}10         & \cellcolor[HTML]{F2F2F2}2          & \cellcolor[HTML]{F2F2F2}0       & \cellcolor[HTML]{F2F2F2}0          & \cellcolor[HTML]{F2F2F2}0      
\end{tabular}
\end{adjustbox}
\end{table}

\subsubsection{\textbf{General programming errors}} This category consists of 5 subcategories and 25 types of causes. The leading cause subcategories are \textit{erroneous method definition and execution}, \textit{incorrect naming and types}, and \textit{compile time error} (see sheet Causes for All Issues in dataset \cite{replpack}). Table \ref{tab:issues-causes} shows the distribution of leading issue categories and GPE causes. Our findings reveal that most GPE causes have become the source for \textbf{code debt}, \textbf{service communication}, and \textbf{build error} issues. We identified 158 causes that become the source for different types of \textit{code debt} issues. %For instance, we identified 28 issues where \textsc{code refactoring} issue due to \textsc{lack of cohesion in methods}. Similarly, we recognized 19 \textsc{code smells} that were due to \textsc{long message chains}. 
We also found several types of causes that become the reasons for more than one issue. For example, \textbf{compile time errors} (e.g., \textsc{semantics errors}) become the reasons for \textit{service communication}, \textit{build errors}, and \textit{broken and missing artifacts} issues.

\textbf{\underline{Finding 7}}: \textbf{General programming errors} identified as the major cause category for microservices issues. \textit{Erroneous method definition and execution}, and \textit{incorrect naming and types}, and \textit{compile time error} found as the major subcategories.%, and \textsc{semantic errors}, \textsc{lack of cohesion in methods}, and \textsc{wrong parameterization} are recognized as the leading types of causes of the issues.

\subsubsection{\textbf{Poor security management}}  
This category consists of 3 subcategories and 17 types of causes mainly related to implementation of security control and policy at \textit{coding}, \textit{communication }, and \textit{application} levels (see sheet Causes for Leading Issues in the dataset \cite{replpack}). Our results show that most of PSM causes have become the source for \textit{authentication and authorization}, \textit{access control}, and \textit{security certificates and standard} issues. For instance, we identified several \textit{authentication and authorization} issues that occur because of writing \textsc{unsafe code}, using \textsc{expired authentication token}, and providing \textsc{malformed input}.   
%(e.g., “\textit{Ädding security plugin with the bug, \#468}”), \textsc{expired authentication token} (e.g., \textit{ “after running the application for some time identity server token expires the frontend,\#472” }), and \textsc{malformed input} (e.g., “Enums or regex validations containing exposed values, \#462”).    

\textbf{\underline{Finding 8}}: Causes appearing in \textbf{poor security management} are account for 87 issues, all belong to \textit{coding} (e.g., \textsc{unsafe code}, \textit{communication} (e.g., \textsc{decryption failure}), and \textit{application} (e.g., \textsc{missing security features}) level subcategories.
\subsubsection{\textbf{Invalid configuration and communication}} We identified 2 subcategories (i.e., \textit{configuration problems}, \textit{communication problems}) and 13 types of causes that become the source for 87 microservices issues. \textsc{Incorrect configuration}, \textsc{service registry errors}, and \textsc{incorrect request handling} are recognized as leading \textbf{ICC} causes (see sheet Causes for Leading Issues in the dataset \cite{replpack}). Our result shows that most \textbf{ICC} causes have become the source for issues from \textit{service communication}, \textit{service execution}, \textit{access control}, and \textit{build error} subcategories (see Table \ref{tab:issues-causes}). For instance, we identified several \textit{service communication} issues  that ensued \textsc{incorrect configuration}, \textsc{limitation of communication protocols}, and \textsc{deprecated localhost}. Similarly, we also find causes to \textit{(build error)} that occurred because of \textsc{incorrect configuration} and \textsc{firewall blocking}. For example, \textsc{Docker build failure} due to \textsc{incorrect configuration}.

%For instance, we identified several \textit{Service communication} issues (e.g., “\textit{Clicking on the login page, eShopOnContainers did not jump to the login page, \#603}”) that ensued \textsc{incorrect configuration} (e.g., “\textit{Wrong OIDC configuration, \#603}”), \textsc{limitation of communication protocols} (e.g., “\textit{Limited support of HTTP/2 for multiplex, \#642}”), and \textsc{deprecated localhost} (e.g., “\textit{Localhost page not found, \#644}”). Similarly, we also find causes to \textit{(Build error)} that occurred because of \textsc{incorrect configuration} and \textsc{firewall blocking}. For instance, \textsc{Docker build failure} due to \textsc{incorrect configuration} (e.g., “\textit{Not passing the registry name to Docker-compose build and push steps, \#342}”) of containers.
\textbf{\underline{Finding 9}}: The types of causes (e.g., \textsc{incorrect configuration}) from the \textbf{invalid configuration and communication} category raise most issues in the leading issue categories (e.g., \textit{service communication}, \textit{build error}, \textit{access control}).

\subsubsection{\textbf{Legacy versions, compatibility, and dependency}}
This category covers the causes for issues that occur when developers use old versions of the software. \textbf{LCD} category consists of 3 subcategories (i.e., \textit{development and deployment}, \textit{client and server}, \textit{compatibility and dependency}), and 14 types of causes that become the source for 49 microservices issues of top four categories. For example, several \textit{build errors} occur due to \textsc{old dev branch}, and \textsc{Kubernetes version}. Similarly, causes (e.g., \textsc{compatibility problem}) from the \textit{compatibility and dependency} subcategory also become the reason of \textit{build error}, \textit{authorization and authentication}, and \textit{secure certificate and connection} issues.
%This category covers the causes for issues that occur when developers use old versions of the software. \textbf{LCD} category consists of 3 subcategories (i.e., \textit{Development and deployment}, \textit{Client and server}, \textit{Compatibility and dependency}) and 14 types (e.g., \textsc{old Dev branch}, \textsc{old Kubernetes version}, \textsc{compatibility problem}) that become the source for 49 microservices issues of top four categories. For instance, several \textit{Build error} (e.g., “\textit{Unable to build eShopOnContainers, \#368}”) occur due to \textsc{old dev branch} (e.g., \textit{ “Using commit from the old DEV branch, \#362}”), and \textsc{Kubernetes version} (e.g., “\textit{Linux K8s setup seems to be out of date, \#413}”). Similarly, causes (e.g., \textsc{compatibility problem}) from the \textit{Compatibility and dependency} subcategory also become the reason of \textit{Build error} (e.g., \textit {“Error NU1202 Package Newtonsoft.Json 11.0.2 is not compatible with netcoreapp2.0}”), \textit{Authorization and authentication} (e.g., “\textit{Can not serve WebSocket using basic auth, \#469}”), and \textit{Secure certificate and connection} (e.g., “\textit{JWT security incompatible with Auth0, \#566}”) issues.

\textbf{\underline{Finding 10}}: 8.4\% (49 out of 584) of identified causes are related to the use of the old version of the software (e.g., identity server), languages (e.g., Go), and deployment platforms (e.g., Kubernetes) that give rise to \textit{build error}, \textit{service communication}, and \textit{access control} issues in open source microservices systems.

%% file: discussion.tex
\section{Discussion}
\label{sec:discussion}
%This section discusses the top four categories of issues in open source microservices systems, their causes and the implications for research and practice.
%We now discuss (i) the top four categories of issues, (ii) their causes, along with (iii) implications of key findings for research and practices, each detailed below.
We now discuss (i) the top four categories of issues, (ii) their causes, along with (iii) implications for research and practices.

%\subsection{Technical Debt Issues, Causes, and Solutions}
%\textbf{Technical debt issues, causes, and solutions.} 

\textbf{Management of TD in microservices systems}:
%Technical debt reflects technical compromises between short-term benefit and long-term health of software systems~\cite{li2015systematic}. 
%Our results indicate that TD demonstrates a wide range of issues related to various facets of open source microservices systems. 
Around one-fourth (23.62\%) of the issues are related to TD, and the signs of the TD can be found in various activities of microservices system development (e.g., design, coding, refactoring, configuration). Our findings reveal that the primary causes behind TD issues are several types of General programming errors, such as lack of cohesion, long message chains, and semantic errors. We observed that TD issues in microservices systems are growing at a higher rate than other types of issues identified in this study. For instance, issues identified for other categories are far less than TD (see Figure \ref{fig:Taxonomy}). Several studies (e.g., \cite {13-taibi2018definition, de2019architectural}) have confirmed the existence of TD in microservices systems from different perspectives (e.g., architectural TD, code smells).
This study results (i.e., the taxonomy of issues, causes of issues) can raise awareness of architects and developers to avoid the accumulation of TD issues before becoming too costly). Despite this, we assert that it could be valuable to investigate some other aspects of TD in microservices systems, such as i) how to design microservices systems to control TD, ii) how TD grows at different levels of microservices systems over time (e.g., design, code, communication), and iii) tools and techniques used to identify, measure, prioritize, monitor, and prevent TD in microservices systems.

\textbf{Build process in microservices systems}:
The build process consists of several activities depending on programming languages, build scripting languages, operating systems, and development processes (e.g., CI/CD, DevOps).10.78\% of the issues are related to the microservices systems' build process, mainly due to missing features and artifacts, legacy versions, compatibility and dependency, and insufficient resources (see Fig. \ref{fig:Taxonomy} and Table \ref{tab:issues-causes}). The build issues and their causes indicate that most build problems occur in issue-triggering phases (e.g., parsing) due to developers' mistakes (e.g., adding dependencies) in build script file. The types of build issues and their causes identified in our study can help practitioners to avoid various types of build issues. For instance, practitioners might avoid using unnecessary dependencies in Docker build files and older versions of Kubernetes while establishing the build and deployment pipeline for microservices systems. Many studies (e.g., \cite{lou2020understanding}) investigated the build issues from monolithic systems in which they claim that developers make non-trivial efforts to address the build issues. However, we did not find any literature that has attempted to investigate what efforts are required to fix the build issues in microservices systems. It would be interesting to compare the types of issues and the required efforts and cost to fix the build issues in microservices systems with monolithic systems. The findings of such studies can help software organizations in the migration of monolithic systems to microservices systems.
%However, we \underline{did not find any literature} in which researchers have investigated what efforts are required to fix build issues in microservices systems. As further research, it can be interesting to compare the type of issues, required efforts, and required cost to fix build issues of microservices systems with monolithic systems, which can further help in the migration of monolithic systems.}

\textbf{Securing microservices systems}: The distributed nature of microservices systems makes them a potential target for cyber-attacks. 10.18\% of the issues belong to the security issue category, mostly because of poor security management of coding, communication, and application-level (see the dataset \cite{replpack}). The issues related to JWT, TSL, SSL, and HTTP cookies type indicate that most of the security issues occurred when microservices are communicating. Similarly, identified causes are also pointing that the security of microservices is poorly managed on a communication level. These results confirm that microservices systems have a much larger attack surface area than traditional systems. The identified security issues and their causes can help practitioners to understand better why and where specific security issues occur in microservices systems. For instance, practitioners might want to avoid writing unsafe code to prevent access control issues. Our findings suggest that security issues are multi-faceted, meaning that security problems can be raised at different levels of microservices systems, such as data centers, cloud providers, virtualization, communication, and orchestration. Therefore, it is valuable to i) explore security valunaribality and related risks at various levels, and ii) propose multi-layered security solutions for fine-grained security management.

\textbf{Executing and communicating microservices}: Microservices systems are distributed systems running on multiple servers or hosts. Microservices communicate through inter-process communication protocols, such as HTTP and AMQ, depending on the type of microservices. We identified that 8.87\% of the issues interrupt execution and communication of microservices mainly because of incorrect configuration and general programming errors. This category of issues and their causes can help practitioners to i) identify the problem areas and ii) make strategies for avoiding the problems of execution and communication of microservices systems (see the dataset \cite{replpack}). Microservices systems may have hundreds of instances that frequently communicate with each other. Service execution and communication of microservices systems can also exacerbate the issue of resiliency, load balancing, distributed tracing, high coupling, and complexity. With these issues in mind, researchers and practitioners can i) propose design techniques for microservices architecture with a particular focus on highly resilient and low coupled microservices systems and ii) develop effective solutions to trace and isolate communication faults.

%% file: threats.tex
\section{Threats to Validity}
\label{sec:threats}
%We discuss two main types of threats, referred to as threats to internal and external validity of this study. %We pinpoint specific methodological steps that we adopted an attempt to minimize the threats, as detailed below.
\textbf{Internal validity} corresponds to methodological rigor to ensure minimal bias in data collection and, ultimately, its impact on study results. As per the methodology in Fig.~\ref{fig:researchmethod}, potential threats to internal validity can be (i) improper selection of OSS projects and data collection and (ii) researchers’ bias in extracting and synthesizing the qualitative data. To minimize internal validity threats, we adopted a multi-step approach to minimize the possible threat stemming from subject system selection (Phase I in Fig.~\ref{fig:researchmethod}). A step-wise and criteria-driven approach for the project selection helped us identify the most appropriate projects (see Table~\ref{tab:selectedProjects}) and eliminate several false positives that may impact internal validity and its impacts on the study results. To mitigate the risk of researchers’ bias in extracting and synthesizing the qualitative data (Phase III, Fig.~\ref{fig:researchmethod}), we defined the explicit criteria (e.g., excluding all those issues that consist of general questions), designed the data extraction forms (see Table \ref{dataitems}, and applied the thematic analysis approach by involving all the authors. Any disagreements about the results of screening, data extraction, and data synthesis process were discussed among all the authors to get a consensus for its resolution. %However, we acknowledge that the five studied projects cannot be representative of all OSS MSA projects in GitHub, and we may have missed some important OSS MSA projects.

\textbf{External validity} of the research refers to the generalization and applicability of the results beyond this study. To maximize the external validity, we started with a large number of issues (2,942 issues) and systematically selected 1,345 issues from the five microservices OSS projects. Besides, to ensure the presentation (types of identified issues) and hierarchical organization of the taxonomy (i.e., types, subcategories, categories), we created an initial version of the taxonomy for evaluation by domain experts. A total of four experts (OSS and MSA researchers and developers) provided us with their feedback to refine and finalize the taxonomy (Fig.~\ref{fig:Taxonomy}). However, the constructed taxonomy of \textit{issues} (RQ1) and the identified categories of \textit{causes} (RQ2) are exclusive to GitHub and may not be comprehensive. We acknowledge that analyzing other open-source microservices systems or proprietary microservice systems may lead to different and/or more comprehensive taxonomies of \textit{issues} and \textit{causes}. We also acknowledge that the five studied projects cannot represent all OSS MSA projects in GitHub, and we may have missed some important open source microservices projects.
Moreover, we provided the dataset of this study (containing issues and causes) online to verify and replicate this study \cite{replpack}.

%\textbf{Construct validity}

%% file: conclusions.tex
\section{Conclusions}
\label{sec:conclusions}
In this work, we conducted an empirical study on the nature of issues (e.g., bugs, faults, errors, failures, mistakes) in open-source microservices systems with a taxonomy of issues based on a manual analysis of 1,345 issue discussions from five OSS projects. The taxonomy is composed of 17 categories containing 46 subcategories and 137 types of issues. We identified that the leading issues stem from technical debt, build, security, and service execution and communication. We also identified and classified the causes for the issues. Overall, this study equips researchers and practitioners with a deep insight into the development and deployment of open source microservices systems. 

As part of future work, we aim for (1) validation of the proposed taxonomy of issues using an industrial survey as well as their potential causes and solutions from the practitioners' perspective, and (2) investigation of the difficulty and priority levels of the identified issues when dealing with them in practice.
%In this work, we present an empirical study on the nature of issues in open source microservices systems.  We have constructed a taxonomy of MSA-based OSS system issues based on manual analysis of 1,345 GitHub. The taxonomy is composed of 17 main categories containing 46 subcategories and 137 unique types of issues. We identified that open source microservices systems’ leading issues are technical debt, build, Security, and service execution and communication. Through this study, we also identified and classified the causes and solutions for microservices system issues. Overall, this study enabled us to get more insights into the issues, causes, and solutions that the developers face during the development of open source microservices systems.
%We observed that developers mainly do not explicitly report the causes, and most of the issues are addressed by inviting pull requests.

%% file: main.bbl

\begin{thebibliography}{35}

%%% the Web address.

\ifx \showCODEN    \undefined \def \showCODEN     #1{\unskip}     \fi
\ifx \showDOI      \undefined \def \showDOI       #1{#1}\fi
\ifx \showISBNx    \undefined \def \showISBNx     #1{\unskip}     \fi
\ifx \showISBNxiii \undefined \def \showISBNxiii  #1{\unskip}     \fi
\ifx \showISSN     \undefined \def \showISSN      #1{\unskip}     \fi
\ifx \showLCCN     \undefined \def \showLCCN      #1{\unskip}     \fi
\ifx \shownote     \undefined \def \shownote      #1{#1}          \fi
\ifx \showarticletitle \undefined \def \showarticletitle #1{#1}   \fi
\ifx \showURL      \undefined \def \showURL       {\relax}        \fi
% invisible to TeX
\providecommand\bibfield[2]{#2}
\providecommand\bibinfo[2]{#2}
\providecommand\natexlab[1]{#1}
\providecommand\showeprint[2][]{arXiv:#2}

\bibitem[\protect\citeauthoryear{Balalaie, Heydarnoori, Jamshidi, Tamburri, and
  Lynn}{Balalaie et~al\mbox{.}}{2018}]%
        {10-balalaie2018microservices}
\bibfield{author}{\bibinfo{person}{A. Balalaie}, \bibinfo{person}{A.
  Heydarnoori}, \bibinfo{person}{P. Jamshidi}, \bibinfo{person}{D. Tamburri},
  {and} \bibinfo{person}{T. Lynn}.} \bibinfo{year}{2018}\natexlab{}.
\newblock \showarticletitle{Microservices migration patterns}.
\newblock \bibinfo{journal}{\emph{Software: Practice and Experience}}
  \bibinfo{volume}{48}, \bibinfo{number}{11} (\bibinfo{year}{2018}),
  \bibinfo{pages}{2019--2042}.
\newblock


\bibitem[\protect\citeauthoryear{Bandeira, Medeiros, Paixao, and Maia}{Bandeira
  et~al\mbox{.}}{2019}]%
        {15-bandeira2019we}
\bibfield{author}{\bibinfo{person}{A. Bandeira}, \bibinfo{person}{C.~A.
  Medeiros}, \bibinfo{person}{M. Paixao}, {and} \bibinfo{person}{P.~H. Maia}.}
  \bibinfo{year}{2019}\natexlab{}.
\newblock \showarticletitle{We need to talk about microservices: An analysis
  from the discussions on StackOverflow}. In \bibinfo{booktitle}{\emph{Proc. of
  the 16th Int. Conf. on Mining Software Repositories (MSR)}}. IEEE,
  \bibinfo{pages}{255--259}.
\newblock


\bibitem[\protect\citeauthoryear{Bogner, Fritzsch, Wagner, and
  Zimmermann}{Bogner et~al\mbox{.}}{2019}]%
        {4-bogner2019assuring}
\bibfield{author}{\bibinfo{person}{J. Bogner}, \bibinfo{person}{J. Fritzsch},
  \bibinfo{person}{S. Wagner}, {and} \bibinfo{person}{A. Zimmermann}.}
  \bibinfo{year}{2019}\natexlab{}.
\newblock \showarticletitle{Assuring the evolvability of microservices:
  Insights into industry practices and challenges}. In
  \bibinfo{booktitle}{\emph{Proc. of the 35th IEEE Int. Conf. on Software
  Maintenance and Evolution (ICSME)}}. IEEE.
\newblock


\bibitem[\protect\citeauthoryear{Borges and Valente}{Borges and
  Valente}{2018}]%
        {Star-GitHub2018}
\bibfield{author}{\bibinfo{person}{H. Borges} {and} \bibinfo{person}{M.~T.
  Valente}.} \bibinfo{year}{2018}\natexlab{}.
\newblock \showarticletitle{What’s in a {GitHub} star? Understanding
  repository starring practices in a social coding platform}.
\newblock \bibinfo{journal}{\emph{Journal of Systems and Software}}
  \bibinfo{volume}{146} (\bibinfo{year}{2018}), \bibinfo{pages}{112--129}.
\newblock


\bibitem[\protect\citeauthoryear{Brings, Daun, Kempe, and Weyer}{Brings
  et~al\mbox{.}}{2018}]%
        {SystematicSearchMap2018}
\bibfield{author}{\bibinfo{person}{J. Brings}, \bibinfo{person}{M. Daun},
  \bibinfo{person}{M. Kempe}, {and} \bibinfo{person}{T. Weyer}.}
  \bibinfo{year}{2018}\natexlab{}.
\newblock \showarticletitle{On different search methods for systematic
  literature reviews and maps: Experiences from a literature search on
  validation and verification of emergent behavior}. In
  \bibinfo{booktitle}{\emph{Proc. of the 22nd Int. Conf. on Evaluation and
  Assessment in Software Engineering (EASE)}}. ACM, \bibinfo{pages}{35--45}.
\newblock


\bibitem[\protect\citeauthoryear{Cito, Schermann, Wittern, Leitner, Zumberi,
  and Gall}{Cito et~al\mbox{.}}{2017}]%
        {2-cito2017empirical}
\bibfield{author}{\bibinfo{person}{J. Cito}, \bibinfo{person}{G. Schermann},
  \bibinfo{person}{J. Wittern}, \bibinfo{person}{P. Leitner},
  \bibinfo{person}{S. Zumberi}, {and} \bibinfo{person}{H.~C. Gall}.}
  \bibinfo{year}{2017}\natexlab{}.
\newblock \showarticletitle{An empirical analysis of the docker container
  ecosystem on {GitHub}}. In \bibinfo{booktitle}{\emph{Proc. of the 14th Int.
  Conf. on Mining Software Repositories (MSR)}}. IEEE,
  \bibinfo{pages}{323--333}.
\newblock


\bibitem[\protect\citeauthoryear{Combe, Martin, and Di~Pietro}{Combe
  et~al\mbox{.}}{2016}]%
        {combe2016docker}
\bibfield{author}{\bibinfo{person}{T. Combe}, \bibinfo{person}{A. Martin},
  {and} \bibinfo{person}{R. Di~Pietro}.} \bibinfo{year}{2016}\natexlab{}.
\newblock \showarticletitle{To docker or not to docker: A security
  perspective}.
\newblock \bibinfo{journal}{\emph{IEEE Cloud Computing}} \bibinfo{volume}{3},
  \bibinfo{number}{5} (\bibinfo{year}{2016}), \bibinfo{pages}{54--62}.
\newblock


\bibitem[\protect\citeauthoryear{Cruzes and Dyba}{Cruzes and Dyba}{2011}]%
        {cruzes2011recommended}
\bibfield{author}{\bibinfo{person}{D.~S. Cruzes} {and} \bibinfo{person}{T.
  Dyba}.} \bibinfo{year}{2011}\natexlab{}.
\newblock \showarticletitle{Recommended steps for thematic synthesis in
  software engineering}. In \bibinfo{booktitle}{\emph{Proc. of the 5th ACM/IEEE
  Int. Symp. on Empirical Software Engineering and Measurement (ESEM)}}. IEEE,
  \bibinfo{pages}{275--284}.
\newblock


\bibitem[\protect\citeauthoryear{de~la Torre, Wagner, and Rousos}{de~la Torre
  et~al\mbox{.}}{2020}]%
        {wagner2018net}
\bibfield{author}{\bibinfo{person}{C. de~la Torre}, \bibinfo{person}{B.
  Wagner}, {and} \bibinfo{person}{M. Rousos}.} \bibinfo{year}{2020}\natexlab{}.
\newblock \bibinfo{booktitle}{\emph{.NET Microservices: Architecture for
  Containerized .NET Applications}}.
\newblock \bibinfo{publisher}{Microsoft Corporation}.
\newblock


\bibitem[\protect\citeauthoryear{De~Toledo, Martini, Przybyszewska, and
  Sj{\o}berg}{De~Toledo et~al\mbox{.}}{2019}]%
        {de2019architectural}
\bibfield{author}{\bibinfo{person}{S.~S. De~Toledo}, \bibinfo{person}{A.
  Martini}, \bibinfo{person}{A. Przybyszewska}, {and} \bibinfo{person}{D.I.
  Sj{\o}berg}.} \bibinfo{year}{2019}\natexlab{}.
\newblock \showarticletitle{Architectural technical debt in microservices: A
  case study in a large company}. In \bibinfo{booktitle}{\emph{Proc. of the 2nd
  Int. Conf. on Technical Debt (TechDebt)}}. IEEE, \bibinfo{pages}{78--87}.
\newblock


\bibitem[\protect\citeauthoryear{Di~Rocco, Di~Ruscio, Di~Sipio, Nguyen, and
  Rubei}{Di~Rocco et~al\mbox{.}}{2020}]%
        {GitHubFilter2020}
\bibfield{author}{\bibinfo{person}{J. Di~Rocco}, \bibinfo{person}{D.
  Di~Ruscio}, \bibinfo{person}{C. Di~Sipio}, \bibinfo{person}{P. Nguyen}, {and}
  \bibinfo{person}{R. Rubei}.} \bibinfo{year}{2020}\natexlab{}.
\newblock \showarticletitle{TopFilter: An approach to recommend relevant
  {GitHub} topics}. In \bibinfo{booktitle}{\emph{Proc. of the 14th ACM/IEEE
  Int. Symp. on Empirical Software Engineering and Measurement (ESEM)}}. ACM,
  \bibinfo{pages}{1--11}.
\newblock


\bibitem[\protect\citeauthoryear{Dragoni, Giallorenzo, Lafuente, Mazzara,
  Montesi, Mustafin, and Safina}{Dragoni et~al\mbox{.}}{2017}]%
        {dragoni2017microservices}
\bibfield{author}{\bibinfo{person}{N. Dragoni}, \bibinfo{person}{S.
  Giallorenzo}, \bibinfo{person}{A.~L. Lafuente}, \bibinfo{person}{M. Mazzara},
  \bibinfo{person}{F. Montesi}, \bibinfo{person}{R. Mustafin}, {and}
  \bibinfo{person}{L. Safina}.} \bibinfo{year}{2017}\natexlab{}.
\newblock \showarticletitle{Microservices: Yesterday, Today, and Tomorrow}.
\newblock In \bibinfo{booktitle}{\emph{Present and Ulterior Software
  Engineering}}. \bibinfo{publisher}{Springer}, \bibinfo{pages}{195--216}.
\newblock


\bibitem[\protect\citeauthoryear{Esposito, Castiglione, and Choo}{Esposito
  et~al\mbox{.}}{2016}]%
        {esposito2016challenges}
\bibfield{author}{\bibinfo{person}{C. Esposito}, \bibinfo{person}{A.
  Castiglione}, {and} \bibinfo{person}{K. Choo}.}
  \bibinfo{year}{2016}\natexlab{}.
\newblock \showarticletitle{Challenges in delivering software in the cloud as
  microservices}.
\newblock \bibinfo{journal}{\emph{IEEE Cloud Computing}} \bibinfo{volume}{3},
  \bibinfo{number}{5} (\bibinfo{year}{2016}), \bibinfo{pages}{10--14}.
\newblock


\bibitem[\protect\citeauthoryear{Fowler and Lewis}{Fowler and Lewis}{2014}]%
        {fowler2014microservices}
\bibfield{author}{\bibinfo{person}{M. Fowler} {and} \bibinfo{person}{J.
  Lewis}.} \bibinfo{year}{2014}\natexlab{}.
\newblock \bibinfo{booktitle}{\emph{Microservices: A definition of this new
  architectural term}}.
\newblock
\urldef\tempurl%
\url{http://martinfowler.com/articles/microservices.html}
\showURL{%
\tempurl}


\bibitem[\protect\citeauthoryear{Gupta and Palvankar}{Gupta and
  Palvankar}{2020}]%
        {Dinkar2020Pitfalls}
\bibfield{author}{\bibinfo{person}{D. Gupta} {and} \bibinfo{person}{M.
  Palvankar}.} \bibinfo{year}{2020}\natexlab{}.
\newblock \bibinfo{booktitle}{\emph{Pitfalls \& challenges faced during a
  microservices architecture implementation}}.
\newblock
\urldef\tempurl%
\url{https://tinyurl.com/49n54h4p}
\showURL{%
\tempurl}


\bibitem[\protect\citeauthoryear{Jamshidi, Pahl, Mendon{\c{c}}a, Lewis, and
  Tilkov}{Jamshidi et~al\mbox{.}}{2018}]%
        {jamshidi2018microservices}
\bibfield{author}{\bibinfo{person}{P. Jamshidi}, \bibinfo{person}{C. Pahl},
  \bibinfo{person}{N.~C. Mendon{\c{c}}a}, \bibinfo{person}{J. Lewis}, {and}
  \bibinfo{person}{S. Tilkov}.} \bibinfo{year}{2018}\natexlab{}.
\newblock \showarticletitle{Microservices: The journey so far and challenges
  ahead}.
\newblock \bibinfo{journal}{\emph{IEEE Software}} \bibinfo{volume}{35},
  \bibinfo{number}{3} (\bibinfo{year}{2018}), \bibinfo{pages}{24--35}.
\newblock


\bibitem[\protect\citeauthoryear{Li, Avgeriou, and Liang}{Li
  et~al\mbox{.}}{2015}]%
        {li2015systematic}
\bibfield{author}{\bibinfo{person}{Z. Li}, \bibinfo{person}{P. Avgeriou}, {and}
  \bibinfo{person}{P. Liang}.} \bibinfo{year}{2015}\natexlab{}.
\newblock \showarticletitle{A systematic mapping study on technical debt and
  its management}.
\newblock \bibinfo{journal}{\emph{Journal of Systems and Software}}
  \bibinfo{volume}{101} (\bibinfo{year}{2015}), \bibinfo{pages}{193--220}.
\newblock


\bibitem[\protect\citeauthoryear{Lou, Chen, Cao, Hao, and Zhang}{Lou
  et~al\mbox{.}}{2020}]%
        {lou2020understanding}
\bibfield{author}{\bibinfo{person}{Y. Lou}, \bibinfo{person}{Z. Chen},
  \bibinfo{person}{Y. Cao}, \bibinfo{person}{D. Hao}, {and} \bibinfo{person}{L.
  Zhang}.} \bibinfo{year}{2020}\natexlab{}.
\newblock \showarticletitle{Understanding build issue resolution in practice:
  symptoms and fix patterns}. In \bibinfo{booktitle}{\emph{Proc. of the 28th
  ACM Joint Meeting on European Software Engineering Conf. and Symp. on the
  Foundations of Software Engineering (ESEC/FSE)}}. ACM,
  \bibinfo{pages}{617--628}.
\newblock


\bibitem[\protect\citeauthoryear{M{\'a}rquez and Astudillo}{M{\'a}rquez and
  Astudillo}{2018}]%
        {1-marquez2018actual}
\bibfield{author}{\bibinfo{person}{G. M{\'a}rquez} {and} \bibinfo{person}{H.
  Astudillo}.} \bibinfo{year}{2018}\natexlab{}.
\newblock \showarticletitle{Actual use of architectural patterns in
  microservices-based open source projects}. In \bibinfo{booktitle}{\emph{Proc.
  of the 25th Asia-Pacific Software Engineering Conf. (APSEC)}}. IEEE,
  \bibinfo{pages}{31--40}.
\newblock


\bibitem[\protect\citeauthoryear{M{\'a}rquez and Astudillo}{M{\'a}rquez and
  Astudillo}{2019}]%
        {8-marquez2019identifying}
\bibfield{author}{\bibinfo{person}{G. M{\'a}rquez} {and} \bibinfo{person}{H.
  Astudillo}.} \bibinfo{year}{2019}\natexlab{}.
\newblock \showarticletitle{Identifying availability tactics to support
  security architectural design of microservice-based systems}. In
  \bibinfo{booktitle}{\emph{Proc. of the 13th European Conf. on Software
  Architecture (ECSA) Companion}}. Springer, \bibinfo{pages}{123--129}.
\newblock


\bibitem[\protect\citeauthoryear{M{\'a}rquez, Villegas, and
  Astudillo}{M{\'a}rquez et~al\mbox{.}}{2018}]%
        {3-marquez2018empirical}
\bibfield{author}{\bibinfo{person}{G. M{\'a}rquez}, \bibinfo{person}{M.
  Villegas}, {and} \bibinfo{person}{H. Astudillo}.}
  \bibinfo{year}{2018}\natexlab{}.
\newblock \showarticletitle{An empirical study of scalability frameworks in
  open source microservices-based systems}. In \bibinfo{booktitle}{\emph{Proc.
  of the 37th Int. Conf. of the Chilean Computer Science Society (SCCC)}}.
  IEEE, \bibinfo{pages}{1--8}.
\newblock


\bibitem[\protect\citeauthoryear{Muntoni, Soldani, and Brogi}{Muntoni
  et~al\mbox{.}}{2020}]%
        {12-muntonimining}
\bibfield{author}{\bibinfo{person}{G. Muntoni}, \bibinfo{person}{J. Soldani},
  {and} \bibinfo{person}{A. Brogi}.} \bibinfo{year}{2020}\natexlab{}.
\newblock \showarticletitle{Mining the architecture of microservice-based
  applications from their Kubernetes deployment}. In
  \bibinfo{booktitle}{\emph{Proc. of the 16th Int. Workshop on Engineering
  Service-Oriented Applications and Cloud Services (WESOACS)}}. Springer,
  \bibinfo{pages}{1--12}.
\newblock


\bibitem[\protect\citeauthoryear{Newman}{Newman}{2020}]%
        {newman2020building}
\bibfield{author}{\bibinfo{person}{S. Newman}.}
  \bibinfo{year}{2020}\natexlab{}.
\newblock \bibinfo{booktitle}{\emph{Building Microservices: Designing
  Fine-grained Systems} (\bibinfo{edition}{second} ed.)}.
\newblock \bibinfo{publisher}{O'Reilly Media, Inc.}
\newblock


\bibitem[\protect\citeauthoryear{Pigazzini, Fontana, Lenarduzzi, and
  Taibi}{Pigazzini et~al\mbox{.}}{2020}]%
        {14-pigazzini2020towards}
\bibfield{author}{\bibinfo{person}{I. Pigazzini}, \bibinfo{person}{F.~A.
  Fontana}, \bibinfo{person}{V. Lenarduzzi}, {and} \bibinfo{person}{D. Taibi}.}
  \bibinfo{year}{2020}\natexlab{}.
\newblock \showarticletitle{Towards microservice smells detection}. In
  \bibinfo{booktitle}{\emph{Proc. of the 3rd Int. Conf. on Technical Debt
  (TechDebt)}}. ACM, \bibinfo{pages}{92--97}.
\newblock


\bibitem[\protect\citeauthoryear{Ren, Gay, K{\"a}stner, and Jamshidi}{Ren
  et~al\mbox{.}}{2020}]%
        {ren2020understanding}
\bibfield{author}{\bibinfo{person}{Y. Ren}, \bibinfo{person}{G. Gay},
  \bibinfo{person}{C. K{\"a}stner}, {and} \bibinfo{person}{P. Jamshidi}.}
  \bibinfo{year}{2020}\natexlab{}.
\newblock \showarticletitle{Understanding the nature of system-related issues
  in machine learning frameworks: An exploratory study}.
\newblock \bibinfo{journal}{\emph{arXiv}}  \bibinfo{volume}{abs/2005.06091}
  (\bibinfo{year}{2020}).
\newblock


\bibitem[\protect\citeauthoryear{Taibi and Lenarduzzi}{Taibi and
  Lenarduzzi}{2018}]%
        {13-taibi2018definition}
\bibfield{author}{\bibinfo{person}{D. Taibi} {and} \bibinfo{person}{V.
  Lenarduzzi}.} \bibinfo{year}{2018}\natexlab{}.
\newblock \showarticletitle{On the definition of microservice bad smells}.
\newblock \bibinfo{journal}{\emph{IEEE Software}} \bibinfo{volume}{35},
  \bibinfo{number}{3} (\bibinfo{year}{2018}), \bibinfo{pages}{56--62}.
\newblock


\bibitem[\protect\citeauthoryear{Taibi, Lenarduzzi, and Pahl}{Taibi
  et~al\mbox{.}}{2017}]%
        {taibi2017processes}
\bibfield{author}{\bibinfo{person}{D. Taibi}, \bibinfo{person}{V. Lenarduzzi},
  {and} \bibinfo{person}{C. Pahl}.} \bibinfo{year}{2017}\natexlab{}.
\newblock \showarticletitle{Processes, motivations, and issues for migrating to
  microservices architectures: An empirical investigation}.
\newblock \bibinfo{journal}{\emph{IEEE Cloud Computing}} \bibinfo{volume}{4},
  \bibinfo{number}{5} (\bibinfo{year}{2017}), \bibinfo{pages}{22--32}.
\newblock


\bibitem[\protect\citeauthoryear{Walker, Das, and Cerny}{Walker
  et~al\mbox{.}}{2020}]%
        {5-walker2020automated}
\bibfield{author}{\bibinfo{person}{A. Walker}, \bibinfo{person}{D. Das}, {and}
  \bibinfo{person}{T. Cerny}.} \bibinfo{year}{2020}\natexlab{}.
\newblock \showarticletitle{Automated code-smell detection in microservices
  through static analysis: A case study}.
\newblock \bibinfo{journal}{\emph{Applied Sciences}} \bibinfo{volume}{10},
  \bibinfo{number}{21} (\bibinfo{year}{2020}), \bibinfo{pages}{7800}.
\newblock


\bibitem[\protect\citeauthoryear{Waseem, Liang, and Shahin}{Waseem
  et~al\mbox{.}}{2020}]%
        {waseem2020systematic}
\bibfield{author}{\bibinfo{person}{M. Waseem}, \bibinfo{person}{P. Liang},
  {and} \bibinfo{person}{M. Shahin}.} \bibinfo{year}{2020}\natexlab{}.
\newblock \showarticletitle{A systematic mapping study on microservices
  architecture in devops}.
\newblock \bibinfo{journal}{\emph{Journal of Systems and Software}}
  \bibinfo{volume}{170} (\bibinfo{year}{2020}), \bibinfo{pages}{110798}.
\newblock


\bibitem[\protect\citeauthoryear{Waseem, Liang, Shahin, Ahmed, and
  Razaei~Nasab}{Waseem et~al\mbox{.}}{2021}]%
        {replpack}
\bibfield{author}{\bibinfo{person}{M. Waseem}, \bibinfo{person}{P. Liang},
  \bibinfo{person}{M. Shahin}, \bibinfo{person}{A. Ahmed}, {and}
  \bibinfo{person}{A. Razaei~Nasab}.} \bibinfo{year}{2021}\natexlab{}.
\newblock \bibinfo{title}{{Dataset of the Paper ``On the Nature of Issues in
  Five Open Source Microservices Systems: An Empirical Study''}}.
\newblock
\newblock
\urldef\tempurl%
\url{https://doi.org/10.5281/zenodo.4602870}
\showDOI{\tempurl}


\bibitem[\protect\citeauthoryear{Wu, Tordsson, Elmroth, and Kao}{Wu
  et~al\mbox{.}}{2020}]%
        {9-wu}
\bibfield{author}{\bibinfo{person}{L. Wu}, \bibinfo{person}{J. Tordsson},
  \bibinfo{person}{E. Elmroth}, {and} \bibinfo{person}{O. Kao}.}
  \bibinfo{year}{2020}\natexlab{}.
\newblock \showarticletitle{MicroRCA: Root cause localization of performance
  issues in microservices}. In \bibinfo{booktitle}{\emph{Proc. of the IEEE/IFIP
  Network Operations and Management Symp. (NOMS)}}. IEEE,
  \bibinfo{pages}{1--9}.
\newblock


\bibitem[\protect\citeauthoryear{Yarygina and Bagge}{Yarygina and
  Bagge}{2018}]%
        {yarygina2018overcoming}
\bibfield{author}{\bibinfo{person}{T. Yarygina} {and} \bibinfo{person}{A.~H.
  Bagge}.} \bibinfo{year}{2018}\natexlab{}.
\newblock \showarticletitle{Overcoming security challenges in microservice
  architectures}. In \bibinfo{booktitle}{\emph{Proc. of the 12th IEEE Int.
  Conf. on Service-Oriented System Engineering (SOSE)}}. IEEE,
  \bibinfo{pages}{11--20}.
\newblock


\bibitem[\protect\citeauthoryear{Yu, Jin, Zhang, and Zheng}{Yu
  et~al\mbox{.}}{2019}]%
        {yu2019survey}
\bibfield{author}{\bibinfo{person}{D. Yu}, \bibinfo{person}{Y. Jin},
  \bibinfo{person}{Y. Zhang}, {and} \bibinfo{person}{X. Zheng}.}
  \bibinfo{year}{2019}\natexlab{}.
\newblock \showarticletitle{A survey on security issues in services
  communication of Microservices-enabled fog applications}.
\newblock \bibinfo{journal}{\emph{Concurrency and Computation: Practice and
  Experience}} \bibinfo{volume}{31}, \bibinfo{number}{22}
  (\bibinfo{year}{2019}), \bibinfo{pages}{e4436}.
\newblock


\bibitem[\protect\citeauthoryear{Zhou, Peng, Xie, Sun, Ji, Li, and Ding}{Zhou
  et~al\mbox{.}}{2018}]%
        {7-zhou2018fault}
\bibfield{author}{\bibinfo{person}{X. Zhou}, \bibinfo{person}{X. Peng},
  \bibinfo{person}{T. Xie}, \bibinfo{person}{J. Sun}, \bibinfo{person}{C. Ji},
  \bibinfo{person}{W. Li}, {and} \bibinfo{person}{D. Ding}.}
  \bibinfo{year}{2018}\natexlab{}.
\newblock \showarticletitle{Fault analysis and debugging of microservice
  systems: Industrial survey, benchmark system, and empirical study}.
\newblock \bibinfo{journal}{\emph{IEEE Transactions on Software Engineering}}
  (\bibinfo{year}{2018}).
\newblock


\bibitem[\protect\citeauthoryear{Zimmermann}{Zimmermann}{2017}]%
        {zimmermann2017microservices}
\bibfield{author}{\bibinfo{person}{O. Zimmermann}.}
  \bibinfo{year}{2017}\natexlab{}.
\newblock \showarticletitle{Microservices tenets}.
\newblock \bibinfo{journal}{\emph{Computer Science-Research and Development}}
  \bibinfo{volume}{32}, \bibinfo{number}{3-4} (\bibinfo{year}{2017}),
  \bibinfo{pages}{301--310}.
\newblock


\end{thebibliography}
